\documentclass[dvipdfmx]{aastex63}
\received{28 Oct. 2019}
\revised{1 Dec. 2019}
\accepted{11 Dec. 2019}
\submitjournal{ApJL}

\shorttitle{The temperature and density of a Spicule observed with ALMA}
\shortauthors{Shimojo et al.}

\begin{document}

\title{Estimating the temperature and density of a spicule from 100 GHz data obtained with ALMA }

\correspondingauthor{Masumi Shimojo}
\email{masumi.shimojo@nao.ac.jp}

\author[0000-0002-2350-3749]{Masumi Shimojo}
\affiliation{National Astronomical Observatory of Japan, National Institutes of Natural Sciences, Mitaka, Tokyo, 181-8588, Japan}
\affiliation{Department of Astronomical Science, The Graduate University of Advanced Studies, SOKENDAI, Mitaka, Tokyo, 181-8588, Japan}

\author[0000-0002-1021-0322]{Tomoko Kawate}
\affiliation{Institute of Space and Astronautical Science, Japan Aerospace Exploration Agency, 3-1-1 Yoshinodai, Chuo-ku, Sagamihara, Kanagawa, 252-5210, Japan}

\author[0000-0003-3765-1774]{Takenori J. Okamoto}
\affiliation{National Astronomical Observatory of Japan, National Institutes of Natural Sciences, Mitaka, Tokyo, 181-8588, Japan}

\author[0000-0001-5457-4999]{Takaaki Yokoyama}
\affiliation{Department of Earth and Planetary Science, The University of Tokyo, Bunkyo-ku, Tokyo 113-0033, Japan}

\author[0000-0002-6330-3944]{Noriyuki Narukage}
\affiliation{National Astronomical Observatory of Japan, National Institutes of Natural Sciences, Mitaka, Tokyo, 181-8588, Japan}

\author[0000-0003-2991-4159]{Taro Sakao}
\affiliation{Institute of Space and Astronautical Science, Japan Aerospace Exploration Agency, 3-1-1 Yoshinodai, Chuo-ku, Sagamihara, Kanagawa, 252-5210, Japan}
\affiliation{Department of Space and Astronautical Science, The Graduate University of Advanced Studies, SOKENDAI, 3-1-1 Yoshinodai, Chuo-ku, Sagamihara, Kanagawa, 252-5210, Japan}

\author[0000-0002-2464-5212]{Kazumasa Iwai}
\affiliation{Institute for Space-Earth Environmental Research, Nagoya University, Furo-cho, Chikusa-ku, Nagoyam 464-8601, Japan}

\author[0000-0001-5557-2100]{Gregory D. Fleishman}
\affiliation{Physics Department, Center for Solar-Terrestrial Research, New Jersey Institute of Technology, Newark, NJ 07102-1982, USA}

\author{Kazunari Shibata}
\affiliation{Astronomical Observatory, Kyoto University, Kitashirakawa Oiwake-cho, Sakyo-ku, Kyoto 606-8502}



\begin{abstract}
We succeeded in observing two large spicules simultaneously with the Atacama Large Millimeter/submillimeter Array (ALMA), the Interface Region Imaging Spectrograph (IRIS), and the Atmospheric Imaging Assembly (AIA) onboard the Solar Dynamics Observatory. One is a spicule seen in the IRIS Mg II slit-jaw images and AIA 304\AA\ images (MgII/304\AA\ spicule). The other one is a spicule seen in the 100GHz images obtained with ALMA (100GHz spicule). Although the 100GHz spicule overlapped with the MgII/304\AA\ spicule in the early phase, it did not show any corresponding structures in the IRIS Mg II and AIA 304\AA\ images after the early phase. It suggests that the spicules are individual events and do not have a physical relationship. To obtain the physical parameters of the 100GHz spicule, we estimate the optical depths as a function of temperature and density using two different methods. One is using the observed brightness temperature by assuming a filling factor, and the other is using an emission model for the optical depth. As a result of comparing them, the kinetic temperature of the plasma and the number density of ionized hydrogens in the 100GHz spicule are $\sim$6800 K and $\rm 2.2\times10^{10} \ cm^{-3}$. The estimated values can explain the absorbing structure in the 193\AA\ image, which appear as a counterpart of the 100GHz spicule. These results suggest that the 100GHz spicule presented in this paper is classified to a macrospicule without a hot sheath in former terminology.
\end{abstract}



\section{Introduction} \label{sec:intro}

A spicule is one of the building blocks of the solar atmosphere and a key-phenomenon for understanding the heating of corona and chromosphere. It is believed that spicules provide the energy and mass for forming the hot atmospheric layers and solar wind. Observations of spicules on the solar limb have been carried out since the 1870s \citep{1870sepd.book.....S} and visible and ultraviolet chromospheric lines (e.g., H$\rm\alpha$, Ca II K) are used for most observations. Due to the recent chromospheric observations with a high-spatial resolution from solar observing satellites, our knowledge of a spicule is rapidly renewing \citep[e.g.,][]{2007PASJ...59S.655D, 2011ApJ...736L..24O,2012ApJ...750...16Z,2002A&A...384..303P}. On the other hand, it is hard to derive physical parameters of the spicules from theses chromospheric lines because of deviating from the local thermodynamic equilibrium (LTE) for these lines. Some authors \citep[e.g.,][]{1972ARA&A..10...73B,1997A&A...324.1183T,2018SoPh..293...20A} determined temperature and density of spicules and their distribution, but the results of them depend on the complicated forward modeling of the radiation from non-LTE medium.

Theoretical studies of spicules have also been widely performed, such as slow shock model \citep[e.g.][]{1961ApJ...134..347O, 1982SoPh...75...99S, 1982SoPh...78..333S,1982ApJ...257..345H, 1988ApJ...327..950S, 1990ApJ...349..647S, 2006ApJ...647L..73H, 2015ApJ...812L..30I},  Alfven wave model \citep[e.g.][]{1982ApJ...257..345H, 1985ApJ...296..746M, 1998A&A...338..729D, 1999ApJ...514..493K, 2012ApJ...749....8M, 2014MNRAS.440..971M, 2013SSRv..175....1M}. Realistic radiative MHD simulation models have also been developed recently \citep[e.g.,][]{2017Sci...356.1269M, 2017ApJ...848...38I},  though more detailed physical analyses are necessary to understand the basic mechanism of splicule formation even for simulation results (See also \cite{2000SoPh..196...79S} and \cite{2012SSRv..169..181T} for the review of spicule theories.) To confirm whether the physical processes in the simulations occur in the solar atmosphere, comparisons between the simulations and the observations are essential. For this purpose, it is required to derive physical quantities of spicules from the observations, but it is not so easy as mentioned above.

Since the millimeter waves emitted from the chromospheric plasma satisfies the LTE condition, it is relatively easier than other chromospheric lines to derive the physical parameters. On the other hand, achieving a high-spatial resolution with millimeter waves require an interferometer with long baselines, and no one had investigated spicules with these wavelength ranges until recently. The Atacama Large Millimeter/submillimeter Array \citep[ALMA;][]{2009IEEEP..97.1463W} is the largest interferometer in the world for observing astronomical objects with millimeter and submillimeter waves, and started scientific solar observations in 2016. The spatial resolution of ALMA for observing the Sun is $\sim$ 2\arcsec \ with 100 GHz and is useful for investigating spicules. The ALMA data already show spicules and plasmoid eruptions \citep{2017ApJ...841L...5S,2018ApJ...863...96Y,2018A&A...619L...6N}.

We obtained the observing time in ALMA Cycle 4 and succeeded in observing spicules with ALMA as well as the Interface Region Imaging Spectrograph \citep[IRIS;][]{2014SoPh..289.2733D}, and the Atmospheric Imaging Assembly  \citep[AIA;][]{2012SoPh..275...17L} onboard the Solar Dynamics Observatory \citep[SDO;][]{2012SoPh..275....3P} simultaneously. In this paper, we present results of the coordinated observation, estimate the temperature and density of the spicules from the ALMA data, and discuss what phenomena in the past observations correspond to spicules observed at 100 GHz. 

\section{Observations} \label{sec:obs}

The ALMA observatory started offering solar observations with Band 3 and Band 6 at Cycle 4, which is the 5th open-use observing period from October 2016 to September 2017. We succeeded in obtaining the observing time in Cycle 4, and our observation was carried out between 14:22 -- 16:13UT on 26 April 2017. We used the Band3 receiver and the Time Division Mode (TDM) of the correlator. Hence, the frequency of the 1st local oscillator is 100 GHz \citep{2016CyclePG}. 

We synthesized one solar image from the visibility data obtained in each integration set whose duration is 2 seconds in the correlator, i.e., the time cadence of the images is 2 seconds. We used the full spectral range obtained with the TDM to enhance the signal-to-noise ratio in an image. Therefore, a synthesized image shows a distribution of intensity at 100 GHz with a bandwidth of about 8 GHz. To obtain synthesized images with high dynamic range, we carried out the self-calibration only for the phase after the standard calibration described in \cite{2017SoPh..292...87S}. The self-calibration has 5 steps with the different accumulating period to obtain the solution from 10 minutes for the first step to 2 seconds for the last step. The CLEAN deconvolution method is used for the image synthesis with the Clark algorithm and Briggs weighting scheme on CASA 5.4.0 \citep{2007ASPC..376..127M}. During the five-steps self-calibration, we used the CLEAN model created in the previous self-calibration as the initial guess of the synthesis. The field of view of the synthesized images is about 1\arcmin. The sizes of the synthesized beam are 2\arcsec.67 and 1\arcsec.88 along the major and minor axes, respectively. Although the singe-dish observations were carried out with the interferometric observations for obtaining the full-sun images \citep{2017SoPh..292...88W}, we did not combine the synthesized images and full-sun images because it is not established how to deal with the solar limb in the full-sun image for the combining. Hence, in this study, we only deal with relative brightness temperatures. They are roughly the same as the differences from the averaged brightness temperature of the field of view. To remove jitter motion caused by the remaining phase error, we spatially shift the images based on the offsets estimated by the auto-correlation of the images.

The coordinated observation with IRIS was executed during the observing period. IRIS obtained slit-jaw images of Mg II (2796\AA) bands with a cadence and a spatial resolution of 40 s and 0\arcsec.40, respectively. Although the spectra of the Mg II lines are obtained with IRIS during the period, we do not use the spectral data because the slit positions do not cover the phenomena described in the next section. We also use 304\AA\ (He II) and 193\AA\ (Fe XII) band images taken with AIA. They provide context filtergrams with a cadence of 12 s. The spatial resolution is 1\arcsec.5. 

For the co-alignment of the images, we compared between the AIA 304\AA\ images and the IRIS Mg II images, and aligned the Mg II images to AIA 304 \AA\ images. We aligned the ALMA images with AIA ones only by using the coordinate information in the data header because the pointing accuracy of ALMA (within 0\arcsec.6  rms) is sufficient for the current analysis \citep{ALMACycle4TH}.

\section{The spicules observed with ALMA, IRIS, and AIA/SDO} \label{sec:spic}

\begin{figure}[h]
\figurenum{1}
\epsscale{.5}
\plotone{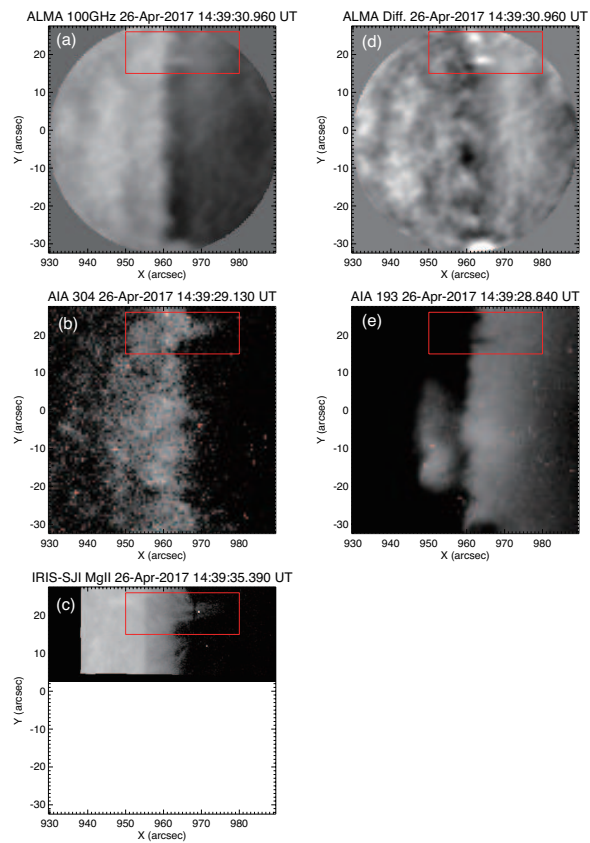}
\caption{Joint observations with ALMA, IRIS, and AIA/SDO. Left column, from top: (a) ALMA 100GHz, (b) AIA 304\AA\ band and (c) IRIS slit-jaw image at Mg II 2796\AA\ band. Right column, from top: (d) ALMA the 100GHz image subtracted the averaged image created from the images obtain within 1.5 hours (ALMA difference image), and (e) AIA 193\AA\ band. The red boxes in the panels indicate the field of view of the panels in Figure 2. The X and Y in the panels present the heliocentric coordinates (SOL-X/Y).\label{fig:all}}
\end{figure}

\begin{figure}[h]
\figurenum{2}
\epsscale{.55}
\plotone{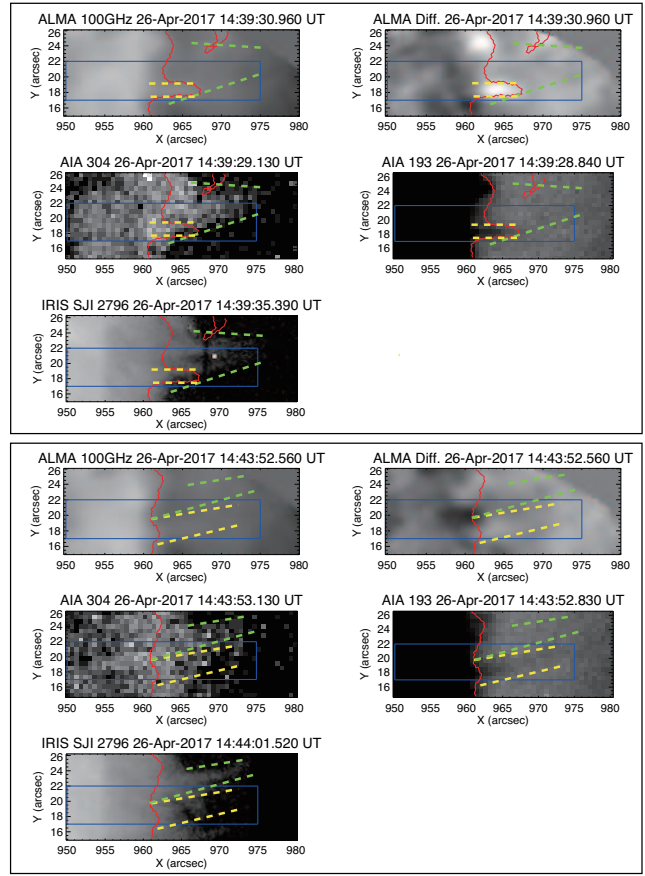}
\caption{The close-up images around the spicules. The order of the panels is the same as that in Figure \ref{fig:all}. Upper panels show the images at 14:39:30UT, and lower panels show the images at 14:43:52UT. The blue boxes show the area for creating the time-slice image (in Figure \ref{fig:ts}). The red contours indicate 50 Jy beam$^{-1}$ ($\sim$170 K) level in the 100GHz images. The green dashed lines indicate the position of the MgII/304\AA\ spicule, and the yellow dashed lines indicate the position of the 100GHz spicule.\label{fig:zoom}\\ 
{\bf Movie 1.} \ The movie covers from 14:35:35UT to 14:45:35UT and the period is repeated twice in the movie. In the second one, the rising phase of the spicule in the Mg II images and 304\AA\ images, and the rising phase of the spicule in the 100GHz and 193\AA\ images are indicated the green, and blue lines, respectively.
\label{fig:video}}
\end{figure}

The panels in Figure \ref{fig:all} show spicules in the 100GHz, 304\AA, 193\AA, and Mg II images. The fields of view of them are the same. In the Mg II image, the solar limb is located around X=956\arcsec, and elongated structures of most spicules cannot be recognized beyond X=963\arcsec. In the 304\AA\ image, we cannot see the counterpart of the photosphere, and the limb at 304 \AA\ is located around the apexes of the spicules seen in the Mg II images. The limb in the 100GHz image is located around X=960\arcsec and is very similar to that in the 193\AA\ image, as reported by \cite{2018ApJ...863...96Y}. We found two large spicules whose apexes are significantly higher than the others seen in the Mg II images. The region that these large spicules appear is indicated by the red boxes in Figure \ref{fig:all}. We concentrate on the phenomena seen in this region and describe them in Figure \ref{fig:zoom} and Movie 1. 

\begin{figure}[h]
\figurenum{3}
\epsscale{.6}
\plotone{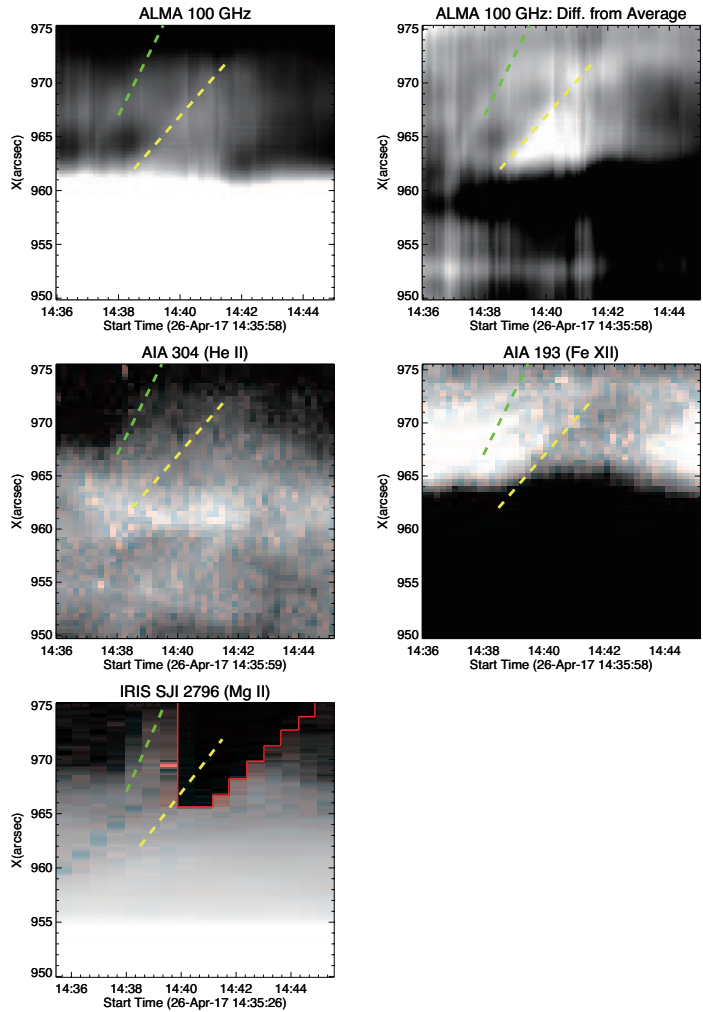}
\caption{The time-slice images created from the data that the areas are indicated by the blue boxes in Figure \ref{fig:zoom}. The order of the panels is the same as that in Figure \ref{fig:all}. The green, and yellow dashed-lines indicate the spicule in AIA 304\AA\ and IRIS 2796\AA\ images, and spicule in ALMA 100GHz and AIA 193\AA\ images, respectively. The slope of the green lines corresponds with 70 km s$^{-1}$, and the slop of the yellow line reveals 40 km s$^{-1}$. The redlines in the IRIS image indicate the area that is not observed. \label{fig:ts}}
\end{figure}

A large spicule appears at around 14:37:50UT above the limb of the Mg II and 304\AA\ images (Figure \ref{fig:zoom} and Movie 1). Because the shapes of the spicules in these images resemble each other, we call the large spicule hereafter ``Mg II/304\AA\ spicule''. In the 304\AA\ images, the Mg II/304\AA\ spicule reached the maximum height at around 14:39UT and the height from the limb seen in the Mg II images are $\sim$20\arcsec ($\sim$15,000 km), and the maximum width is about 4 -- 5\arcsec \ (2,900 -- 3,700 km). The rising velocity of the Mg II/304\AA\ spicule is about 70 km $\rm s^{-1}$ (the green line in Figure  \ref{fig:ts} ). We cannot find the counterpart of the spicule in 100GHz and 193\AA\ images.

About one minute later from the start of the Mg II/304\AA\ spicule (14:38:40UT), another large spicule appeared above the limb in the 100 GHz images (Movie 1). We call the spicule ``100GHz spicule'' in this paper. As pointed out in \cite{2017SoPh..292...87S}, the brightness just above the solar limb in the synthesized images is decreasing gradually with the distance from the limb. The brightness profile does not show the actual brightness profile of the limb that should show steeply decreasing at the limb. It is caused by the lack of baselines. In other words, the number of ALMA's antennas is not enough for synthesizing images of the solar limb. To remove the influence of the artificial structure,  we estimated the brightness temperature of the 100GHz spicule from subtracting the brightness at the pre-event, from the brightness of the events, pixel by pixel. As a result, the peak enhancement of the brightness temperature caused by the spicule is about 240 K. This signal level is much larger than the noise, i.e. $\sim$2.6 K (Shimojo et al. 2017b). An absorbing structure appeared in the 193\AA\ images, and its shape is similar to the spicule in the 100GHz image. The rising velocity of the spicule seen in the 100GHz and 193\AA\ images is about 40 km $\rm s^{-1}$  (the yellow line in Figure \ref{fig:ts}). The 100GHz spicule reached the maximum height of $\sim$15\arcsec\ ($\sim$11,000 km) from the limb seen in the Mg II images at 14:41UT. The obtained width in 100 GHz image is 2\arcsec\ ($\sim$1,500 km), but the actual width can be narrower than the value because the width is comparable to the size of the synthesized beam.

The 100GHz spicule is overlapped by the Mg II/304\AA\ spicule at 14:38:40UT, as shown in the upper panels of Figure \ref{fig:zoom}. Nevertheless, we cannot find any remarkable enhancements in Mg II and 304\AA\ images that correspond to the 100GHz spicule. After 14:43UT (the lower panels in Figure \ref{fig:zoom}), there is no overlapped region between the 100GHz spicule and Mg II/304\AA\ spicule. Moreover, the apparent velocity of the 100 GHz spicule ($\sim$40 km s$^{-1}$) is significantly slower than the Mg II/304A spicule ($\sim$70 km s$^{-1}$), as shown in Figure \ref{fig:ts}. The facts would suggest that the spicules are individual events and do not have a physical relationship. 

\begin{figure}[h]
\figurenum{4}
\epsscale{0.6}
\plotone{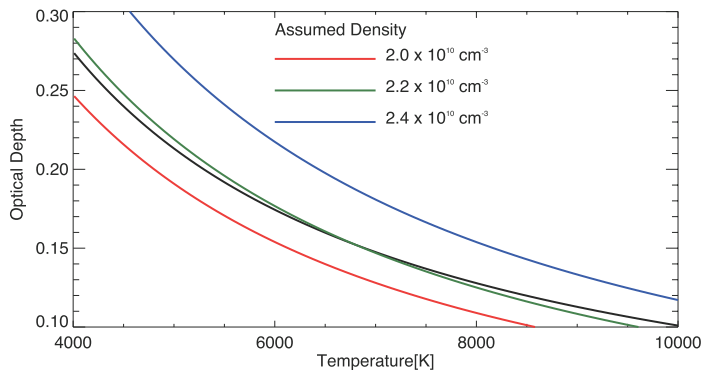}
\caption{Estimated optical depths as a function of assumed temperatures. The solid black line indicates the optical depth at 100 GHz that is estimated from the brightness temperatures considering the filling factor, and assumed temperatures of the spicule. The colored lines reveal the theoretical optical depth at 100 GHz assuming the temperature, density, diameter of the spicule, and the emission via the thermal free-free interaction. The color indicates the difference of the assumed density.\label{fig:od}}
\end{figure}

Considering the typical formation temperatures of the Mg II band \citep[5000 -- 16,000 K:][]{2014SoPh..289.2733D}  and 304\AA\ band \citep[$\sim$ 50,000 K:][]{2012SoPh..275...17L}, the temperature of the 100GHz spicule should be lower than about 10,000 K because we cannot find its counterpart in Mg II and 304\AA\ images. Can the plasma with such temperature and density be explained by the observed structure at 100 GHz? To answer the question, we estimate the optical depths of the spicule at 100 GHz from observations and an emission model, and examine the temperature and density of the 100GHz spicule. Before the estimations, we must consider a filling factor because the spatial resolution of the 100GHz images is not enough for resolving the spicule. The spatial resolution of AIA is not enough either, so that we assume that the width of the 100GHz spicule is 0\arcsec.5. Since the apparent width of the 100GHz spicule is 2\arcsec, the filling factor of the spicule in the 100GHz image is roughly 0.25. Thus, we use 960 K (= 240 K/0.25) as the brightness temperature of the 100GHz spicule. From the observed enhancement of the brightness temperature, the optical depth of the spicule as a function of electron temperature  ($\tau = - log_{e}(1-T_{b}/T)$ $\tau$: optical depth, , $T$: electron temperature, $T_{b}$: brightness temperature) is shown by the black solid line in Figure \ref{fig:od}, based on the following assumptions: 1) the background emission is negligibly small because the phenomenon locates in the off-limb region, 2) the temperature range of the 100GHz spicule is lower than 10,000K, 3) the emission from the spicule is the thermal free-free emission (electron-proton free-free) from the medium satisfied the LTE condition. On the other hand, by assuming the temperature and density of the spicule and the line-of-sight diameter of the spicule is 0\arcsec.5 (same as the width),  we derived the optical depth of the thermal free-free emission from the emission model \citep[e.g.,][]{1980afcp.book.....L,1985ARA&A..23..169D,2018ApJ...863...96Y}. The model functions indicate the colored lines in Figure \ref{fig:od}. The black line estimates from the observations and the colored line of the model cross when we assume the kinetic temperature of the plasma and the number density of ionized hydrogens to be $\sim$6800 K and $\rm2.2\times10^{10} \ cm^{-3}$ respectively and the values indicate the physical parameters of the 100GHz spicule. 

\cite{1998ApJ...504L.127Z} presented that the absorption at the limb in 195\AA\ images show the chromosphere, where the number density of neutral hydrogens is  $\rm 10^{10} cm^{-3}$.  Therefore, for explaining the absorbing counterpart of the 100GHz spicule in the 193\AA\ images, the number density of neutral hydrogens should be similar or larger than the number density of the electrons in the temperature range that the thermal free-free emission is dominant. According to the lower-right panel of Figure 1 in \cite{2017A&A...598A..89R}, the temperature of the 100GHz spicule should be 4000 -- 7000 K. The temperature range is consistent with the temperature derived above, and thus, justifies our estimation. Moreover, the total number density of hydrogens, which is the sum of the number densities of ionized and neutral hydrogens, is assumed $\rm 10^{11} cm^{-3}$ in the lower-right panel of Figure 1 in Rutten (2017). Hence, the mass density of the 100GHz spicule might be about $\rm \sim10^{-13} g \ cm^{-3}$ that is consistent with the previous results of the infrared observations \citep{2000SoPh..196...79S}.

\section{Discussion}\label{sec:disc}

The spicules described in this paper would be categorized into ``macrospicule'' formerly because their sizes and velocities are larger than the typical values of spicules. Therefore, it is appropriate to compare our results with the properties of macrospicules, rather than those of spicules. Macrospicules are most often visible in EUV transition-region lines, such as in He II 304 \AA, and consist of a cool core and hot sheath \citep[e.g.,][]{1975ApJ...197L.133B,1976ApJ...203..528W,2002A&A...384..303P,2019ApJ...871..230L}. Since the Mg II/304\AA\ spicule has a cool component revealed in the Mg II image and a hotter component shown in the He II 304\AA\ images, it has the typical properties of macrospicules described in the preceding papers.

On the other hand, for the 100 GHz spicule, we do not find its counterpart in Mg II/304\AA\ images. \cite{1991ApJ...376L..25H}  presented that the brightness temperature of macrospicules at 4.8, 8.5, and 15 GHz can be explained by an empirical model that macrospicules consist of a cool core at $\sim$8000K surrounding by a hot sheath at $\rm 1 $ -- $\rm 2 \times 10^{5}$ K. The temperature of the cool core is similar to the 100GHz spicule, and there is no significant signal of hotter lines around the 100GHz spicule. Hence, the 100GHz spicule described in this paper is a macrospicule without a hot sheath that is already reported by some authors \citep[e.g.,][]{1975ApJ...197L.133B,2011A&A...532L...1M}. 

The height of the limb in 100GHz images is similar to the average heights of the typical spicules seen in Mg II images, and the limb seen in 304\AA\ images. It suggests that the 100GHz limb is covered by multiple spicules in Mg II and He II images. Hence, we cannot identify a hot sheath around the 100GHz spicules, but we cannot argue that a spicule without a hot sheath is common in 100GHz images. To reveal the properties of typical spicules seen in mm-wave images, we need the higher resolution enough for resolving spicules, which will be realized with the higher observing frequency and/or long baselines of ALMA.

\acknowledgments
This paper makes use of the following ALMA data: ADS/JAO.ALMA \#2016.1.00070.S. ALMA is a partnership of ESO (representing its member states), NSF (USA), and NINS (Japan), together with NRC (Canada), MOST, ASIAA (Taiwan), and KASI (Republic of Korea), in cooperation with the Republic of Chile. The Joint ALMA Observatory is operated by ESO, AUI/NRAO, and NAOJ. IRIS is a NASA small explorer mission developed and operated by LMSAL with mission operations executed at NASA Ames Research Center and major contributions to downlink communications funded by ESA and the Norwegian Space Centre. SDO is part of NASA's Living With a Star Program. The authors are supported by JSPS KAKENHI Grants: M.S. is by JP17K0539, T.K is by JP15H05814 and JP17K14314, T.J.O is by JP16K17663 (PI: T.J.O.) and JP25220703 (PI: S. Tsuneta), T.Y. is by JP15H03640. K.I. is by JP18H04442. G.F. was supported in part by NSF grant AST-1820613 to the New Jersey Institute of Technology. The study was started from the ALMA workshop ``ALMA-Sol-CDAW19'' held in January 2019 that is supported by the ALMA project, NAOJ. 

\facilities{ALMA, SDO, IRIS} 

\bibliography{Alma_Spicule2019}{}
\bibliographystyle{aasjournal}


\end{document}